\documentclass[12pt,preprint]{aastex}
\usepackage{graphics,latexsym,psfig}

\def\etal{{et\,al.}}
\def\msun{M$_{\odot}$}

\def\degs{\ifmmode ^{\circ}\else$^{\circ}$\fi}
\def\amin{\ifmmode ^{\prime}\else$^{\prime}$\fi}
\def\asec{\ifmmode ^{\prime\prime}\else$^{\prime\prime}$\fi}
\def\fdg{\hbox{$.\!\!^\circ$}}          
\newbox\grsign \setbox\grsign=\hbox{$>$}
\newdimen\grdimen \grdimen=\ht\grsign
\newbox\laxbox \newbox\gaxbox
\setbox\gaxbox=\hbox{\raise.5ex\hbox{$>$}\llap
     {\lower.5ex\hbox{$\sim$}}}\ht1=\grdimen\dp1=0pt
\setbox\laxbox=\hbox{\raise.5ex\hbox{$<$}\llap
     {\lower.5ex\hbox{$\sim$}}}\ht2=\grdimen\dp2=0pt
\def\gax{$\mathrel{\copy\gaxbox}$}

\def\chan{{\it Chandra}}

\begin{document}

 \title{Chandra observations of the recurrent nova CI Aql after its 
  April 2000 outburst}

 \author{J. Greiner}
 \affil{Astrophysikalisches Institut
      Potsdam, An der Sternwarte 16, 14482 Potsdam, Germany}
 \affil{Max-Planck-Institut f\"ur extraterrestrische Physik,
      85741 Garching, Germany}
  \email{jcg@mpe.mpg.de}

  \and

  \author{R. Di\,Stefano\altaffilmark{2, 3}}
  \affil{Harvard-Smithsonian Center for Astrophysics, Cambridge, MA 02138}
  \affil{Department of Physics and Astronomy, Tufts University, Medford, MA 02155}
 \email{rdistefano@cfa.harvard.edu}


\date{Version \today ; \hspace{3cm} Received 5 July 2002 / Accepted 4 Sep 2002}

\begin{abstract}
We report the results of two \chan\ observations of the recurrent nova
CI~Aql at 14 and 16 months after its outburst in April 2000, respectively. 
The X-ray emission is faint in both cases, without any noticeable
change in spectrum or intensity. Although the emission is very soft,
it is not luminous enough to be due to 
late-time H-burning. This 
implies that the luminous supersoft phase ended even
before the time predicted by the most recent calculations.
The details of the X-ray spectrum, together with the fact that the
observed X-ray intensity is brighter than pre-outburst (1992/1993),
suggest that the observed X-ray emission is either due to ionization of the 
circumstellar material or due to the shocks within the wind and/or 
with the surrounding medium.
\end{abstract}


\section{Introduction}

Recurrent novae (RN) are a rare subclass of cataclysmic variables,
showing repeated outbursts with typical recurrence times of 50--100 yrs.
These outbursts are thought to be thermonuclear runaways on the
surface of a white dwarf that has accreted enough material from the
companion star to ignite the surface hydrogen.
If the hot white dwarf could be observed
directly, it would have the characteristics of a supersoft X-ray source (SSS);
i.e., its luminosity would lie between $\sim 10^{36}$ ergs/s and $10^{38}$
ergs/s, with $k\, T$ on the order of tens of eV 
(Nomoto 1982, Fujimoto 1982, Iben 1982).
The short recurrence times require either or both small envelope masses and 
high accretion rates of the order of $\sim$10$^{-8}$ \msun/yr (Warner 1995).
Since the ejecta of RN typically are not metal-enriched, the white dwarfs in 
RN are not significantly eroded but may grow in mass (Starrfield \etal\ 1985).
The fitting of RN optical light curves suggests that the white dwarf
masses are high, near the Chandrasekhar limit (e.g. Hachisu \& Kato 1999, 
Hachisu \etal\ 2000), suggesting that RN may be progenitors of type Ia 
supernovae.

Though only 8 other recurrent novae are known, three subclasses are 
defined (Warner 1995)
according to the nature of the companion and, consequently,
the length of orbital period: the T Pyx subclass with dwarf companions
($P \sim 2-3$ hrs), the U Sco subclass with evolved main-sequence or
subgiant companions ($P \sim 1$ day), and the RS Oph subclass with 
red giant companions ($P$ \gax\ 100 days).

CI Aql was discovered when it showed a $\Delta$m = 4.6 mag
outburst in 1917 (Reinmuth 1925). It was classified as a dwarf nova
with long cycle length
(Duerbeck 1987), but this
was cast in doubt (Szkody and Howell 1992, Greiner \etal\ 1996) 
until a second outburst,
discovered in April 2000 (Takamizawa 2000), revealed CI Aql to be a RN.
A subsequent search in the Harvard Plate collection led to the discovery
of another outburst in 1941/1942 (Schaefer 2001b).
CI Aql is an eclipsing binary system with an orbital period of 14.84 hrs
(Mennickent \& Honeycutt 1995) and therefore belongs to the U Sco subclass.

Recurrent novae  are expected to emit soft X-rays during a short
interval after the ejected shell has become optically thin and before the
hydrogen shell-burning ceases and during an interval of cooling.  
During this time, the effective radius should be somewhat larger
than the radius of the white dwarf, and the luminosity should therefore
be larger than 10$^{37}$ erg/s. Thus it should fit the profile of a SSS.
The only previous recurrent novae observed
with X-ray detectors were T Pyx and U Sco. 
T Pyx has had its last outburst in 1967, but was argued to be a 
quasi-persistent supersoft X-ray source because optical data suggested a
luminosity of $\sim$10$^{36}$ erg/s (Patterson \etal\ 1998).
T Pyx was not detected, however, in a ROSAT PSPC
observation carried out in December 1998 (Greiner, unpublished).
A BeppoSAX observation carried out 20 days after the U Sco
outburst revealed a luminous ($\sim$10$^{36}$ erg/s) soft X-ray component
with a best-fit temperature of $\sim$9$\times$10$^5$ K, consistent with
steady nuclear burning continuing for at least $\sim$1 month after
the nova outburst (Kahabka \etal\ 1999).
Further observations of the evolution of the soft and hard X-ray emission 
in other sources are crucial to better test theoretical models 
and also to test the conjecture that RN are SN Ia progenitors. 

Based on the observed optical light curve of CI Aql over the first year
after the 2000 outburst, Hachisu and Kato (2001) predicted that CI Aql should 
have been active at soft X-ray wavelengths between December 2000
until August 2001.  This motivated us to observe CI Aql using \chan.

\section{Observations and Results}

We obtained two \chan\ observations using ACIS-S. 
The first one was performed on June 1, 2001 for 2.15 ksec, and the second
one on August 1, 2001 for a total of 19.88 ksec (see Tab. \ref{obslog}
for details).

During the first \chan\ observation, we detected CI Aql with a count rate
of (6.0$\pm$2.3)$\times$10$^{-3}$ cts/s, collecting a total of 15 photons.
In the August observation, the mean count rate is 
(8.6$\pm$0.7)$\times$10$^{-3}$ cts/s, thus
CI Aql exhibited the same X-ray intensity as on 1 June 2000,

No single-component model fits the X-ray spectrum of CI Aql (second
observation; Figs. 1--3).
As a first approximation, and since the spectrum looks thermal, 
we have used a blackbody spectrum plus a
power law, and find a reasonably good fit with $kT$=50$\pm$20 eV
and a power law photon index of 1.78$\pm$0.72. Note, though, that 
due to the small number of photons the fit parameters are correlated, 
and therefore
the reduced $\chi^2_{\rm red}$ is smaller than one. Using a white dwarf
atmosphere model (van Teeseling \etal\ 1994) gives a similar value 
of $kT$=38$\pm$15 eV.
The best-fit absorbing column is 
3.6$\times$10$^{21}$ cm$^{-2}$ (for the blackbody model)
or 5.5$\times$10$^{21}$ cm$^{-2}$ (for the white dwarf atmosphere model), 
consistent with the
foreground extinction of E(B-V) = 0.85  (Kiss \etal\ 2001),
but considerably less than the total galactic column in this 
BII=--0\fdg8 direction of 1.5$\times$10$^{22}$ cm$^{-2}$
(Dickey \& Lockman 1990). 
The unabsorbed luminosity of the blackbody (white dwarf atmosphere) 
component is 1.0$\times$10$^{33}$ (D/1 kpc)$^2$ erg/s
(8.0$\times$10$^{32}$ (D/1 kpc)$^2$ erg/s).

Physically, such a high-temperature optically thick emission at
such a low luminosity is difficult to explain. The corresponding
radius of the emitting area would be 40 km, a minute fraction
of the white dwarf radius. We therefore turned to other spectral models.

In some recent  X-ray grating observations of novae, the spectra show
a wealth of narrow emission lines in the 0.2--2 keV range. Though most of
the identifications of these lines and their origin are not settled,
we considered the possibility that pure line emission at soft energies
could model the spectrum of CI Aql: 
To model the main emission around 0.5 keV we need only one Gaussian
line with a width corresponding to the energy resolution of ACIS-S.
The best-fit line energy is 0.358 keV.
The spectrum at high energies (above $\sim$4) can similarly be fit by models
other than a power law.
A Raymond-Smith model gives a best-fit temperature of $kT$ = 2.7$\pm$2.4 keV
a thermal bremsstrahlung model gives $kT$ = 3.1$\pm$3.0 keV.
Of course, the spectrum could also be explained by a 
superposition of several lines, but we do not have enough photons
and/or spectral resolution to resolve this issue.

The first-epoch spectrum can be fit by
the model parameters of the fit to the second-epoch observation.
This is not a great surprise given the few photons in the first-epoch 
spectrum.
Yet, we can conclude that 
(1) the intensity was seemingly steady, and 
(2) no major spectral change (change in the emission mechanism) occurred 
between the two \chan\ observations both in the soft as well as in the hard
component --
this also excludes any dramatic decrease of absorption which is expected
once the expelled shell gets optically thin for X-rays,
or cooling of the soft component
(though this cannot be directly proven with the 15 detected X-ray photons).

The second \chan\ observation covers about 40\% of the orbital period
of CI Aql. We therefore analyzed the temporal behaviour of the X-ray 
emission, but did not find orbital variability.

\section{Discussion}

After our \chan\ observations occurred, more optical data became available
(Schaefer 2001a), leading Hachisu \& Kato (2002a) to revise their modelling 
of the optical
decay light curve of CI Aql. The important new result of their revision
 for the interpretation
of our \chan\ data is that the supersoft X-ray phase should have
lasted only until May 2001, thus ending even before the first of our
\chan\ observations. But the source would have remained hot and several orders
of magnitude more luminous than observed had nuclear burning ended as 
recently as May 2001.

The faintness of the detected X-ray emission and the fit results of
a blackbody (or white dwarf atmosphere) model
indeed suggest that CI Aql was not in
the nova plateau phase during our \chan\ observations. 
Instead, the luminosity derived for an
optically thick emission model explaining the soft part of the X-ray spectrum 
is only of the order of 10$^{33}$ (D/1 kpc)$^2$ erg/s. 

However, even beyond the plateau phase the system should have
be observable as a SSS, because the cooling of the hot white dwarf
and the concordant luminosity decay is expected to last months to years.
Model calculations (Hachisu \& Kato 2002b) show that the temperature 
should have declined by a factor 
of $\sim$2 (from 55 eV to 30 eV)  between our two \chan\ observations,
whereas the \chan\ data do not allow a temperature change by more than 10 eV.
These model calculations also suggest that in the June--August 2001
time frame the luminosity of the cooling white dwarf would still be in the 
10$^{37} - 10^{36}$ erg/s range, at least a factor of 1000 more than
the observed X-ray luminosity. 
Thus, we find it very unlikely that the soft X-ray component around
0.5 keV is an optically thick component from the (cooling) white dwarf.
This implies that nuclear burning ended earlier than indicated by even 
the revised prediction.

Alternatively, the soft component could be composed of (a) non-resolved
emission line(s). After fitting a Gaussian line to the soft component,
one realizes that the spectral resolution of $\sim$100 eV makes
this Gaussian line broad enough to fit perfectly the soft component,
and not even a second Gaussian line is necessary to explain the
X-ray emission below 0.7 keV. The best-fit line energy for this
Gaussian is 0.358$\pm$0.058 keV, and the unabsorbed line luminosity
2.7$\times$10$^{32}$ (D/1 kpc)$^2$ erg/s.
We suggest this line to be the H-like C\,VI line (nominal energy at
0.3676 keV). 

Rebinning the spectrum into only 1$\sigma$ confidence level bins 
(Fig. \ref{xsp4sig}),
the search for possible further excess emission above the broad
band continuum emission at higher energies reveals marginal evidence for
possible lines at 0.8 keV, 1.28 keV and 6.3 keV. These could be
interpreted as O\,VII, Ne\,IX or Mg\,X and Fe\,XXVI lines. Given the
signal-to-noise ratio of the spectrum, this is speculative.

The hard spectrum is not constrained: it could be fit by either a power law,
a Raymond-Smith model or a bremsstrahlung model.
Also the marginal Fe detection does not provide a clue for its origin
(e.g. reflection off the disk, or thermal plasma). Indeed, the
Fe emission could be due to matter from the white dwarf, consistent 
with the strong FeII lines seen in the optical spectra (Kiss \etal\ 2001).
The small temperature of order of 3 keV, if interpreted as thermal emission,
is consistent with CI Aql having had a very slow nova outburst, and
a rather low expansion velocity, a correlation which has been found already
in earlier novae (e.g. Mukai \& Ishida 2001 for Nova V382 Vel).

We note, on the other hand, that the unabsorbed luminosity of this hard 
component of $\sim$10$^{30}$ (D/1 kpc)$^2$ erg/s (0.5--10 keV) is in the range
of X-ray luminosities usually found in cataclysmic binaries during
quiescence. However, CI Aql during quiescence is definitely fainter:
ROSAT observations in 1992 and 1993  failed
to detect CI Aql, implying an upper limit on the unabsorbed
X-ray luminosity of 5$\times$10$^{29}$ erg/s (D/1 kpc)$^2$ 
in the 0.1--2.4 keV range
(Greiner \etal\ 1996). Thus, 16 months after its outburst, 
CI Aql had not yet returned to its quiescent level of X-ray emission.

Thus, it seems clear that we observed the CI Aql system when it had ended
both the phase of shell burning and cooling as well as the phase in which the 
wind was obscuring the soft emission. 
In the cases of Nova V382 Velorum (Burwitz \etal\ 2002)
and V1494 Aql (Krautter \etal\ 2002) it has recently been possible to
observe the dramatic changes of the X-ray emission of novae
from a highly absorbed spectrum during the very early times, over
the development and subsequent fading of a luminous supersoft X-ray component
towards a completely line dominated spectrum with hardly any detectable
continuum.
While the first two stages of this evolution are physically well understood
in terms of the expanding shell blocking less and less of the X-ray 
radiation of the white dwarf until burning on its surface ceases,
the origin of the line dominated spectrum is not clear.
One possibility is that during the late phase of a nova,
when the shell has become transparent, the medium around
the system is ionized by the still UV-EUV bright white dwarf.
It would require far less than one part in a thousand conversion efficiency of
photoionization of the C VI line to explain, e.g., the spectrum of CI Aql
as observed with \chan, consistent with photoionization models
around supersoft ionizers (Rappaport \etal\ 1994). However, because of the
steep and rapidly changing density profile due to the expanding shell/wind,
the recombination emission may be rapidly suppressed.
Alternatively, the line dominated spectrum could result from the
interaction of the expanding shell with the circumstellar matter,
shocks within the expanding shell, or from shocks due to collisions
of a fast wind with the interstellar matter. Detailed investigations
of the complex emission line spectra have just begun, and promise
a better understanding in the near future.

\acknowledgements
We are indebted to H. Tananbaum for granting DDT time for the {\chan} 
observations. 
We thank I. Hachizu and M. Kato for prompt information on their modelling
of CI Aql.
This material is based upon work supported by the National Science Foundation
under Grant No. 9815655 and by a NASA LTSA award, NAG5-10705.

\newpage

\begin{table*}
\begin{center}
\caption{Observation log \label{obslog}}
\begin{tabular}{crccc}
  \tableline
   \noalign{\smallskip}
    Obs.-Interval (UT) & Expo.-Time & Seq.-Num. & ObsID & Count rate\\
              & (ksec)~~~     &           &       &  (cts/ksec)  \\
  \noalign{\smallskip}
   \tableline
   \noalign{\smallskip}
     2001-06-01 20:21--21:31 &  2.15~~ & 300054 & 2465 & 6.0$\pm$2.3 \\
     2001-08-01 12:24--18:27 & 19.88~~ & 300056 & 2492 & 8.6$\pm$0.7 \\
  \noalign{\smallskip}
   \tableline
\end{tabular}
\end{center}
\end{table*}

\newpage

\begin{table*}
\begin{center}
\caption{Parameters of the spectral fits.
  $N_{\rm H}$ = absorbing column (in 10$^{21}$ cm$^{-2}$);
  $E_{\rm L}$ = Line energy; $W_{\rm L}$ = Line width.
\label{res}}
\begin{tabular}{ll}
  \tableline
   \noalign{\smallskip}
    Model component & Parameters \\
  \noalign{\smallskip}
   \tableline
  \noalign{\smallskip}
    \multicolumn{2}{l}{Blackbody + power law: $N_{\rm H}$ = 3.6$\pm$0.2;
           $\chi_{\rm red}$=0.69} \\
      ~~blackbody & $kT$ = 50$\pm$20 eV  ~~Norm = 3.18E-05 \\
      ~~power law & $\alpha$ = 1.78$\pm$0.72 ~~Norm = 2.16E-06 \\
   \noalign{\bigskip}
    \multicolumn{2}{l}{White dwarf atmosphere + power law: $N_{\rm H}$ = 5.5$\pm$0.2;
           $\chi_{\rm red}$=0.72} \\
      white dwarf atmosphere & $kT$ = 43$\pm$15 eV  ~~Norm = 1.58E-05 \\
      ~~power law & $\alpha$ = 1.86$\pm$0.95 ~~Norm = 5.42E-06 \\
   \noalign{\bigskip}
    \multicolumn{2}{l}{1 Gaussian + bremsstrahlung: $N_{\rm H}$ = 2.9$\pm$0.2;
           $\chi_{\rm red}$=0.60} \\
      ~~Gaussian 1     & $E_{\rm L}$ = 0.363 ~~$W_{\rm L}$ = 0.11 ~~Norm = 7.03E-04 \\
      ~~Bremsstrahlung & $kT$ = 3.1$\pm$3.0 eV  ~~Norm = 3.25E-06 \\
   \noalign{\bigskip}
    \multicolumn{2}{l}{2 Gaussians + bremsstrahlung: $N_{\rm H}$ = 2.9$\pm$0.2;
           $\chi_{\rm red}$=0.64} \\
      ~~Gaussian 1     & $E_{\rm L}$ = 1.254  ~~Norm = 3.59E-07 \\
      ~~Gaussian 2     & $E_{\rm L}$ = 0.364  ~~Norm = 7.03E-04 \\
      ~~Bremsstrahlung & $kT$ = 6.1$\pm$3.0 eV  ~~Norm = 2.45E-06 \\
   \noalign{\bigskip}
   \tableline
\end{tabular}
\end{center}
\end{table*}

\newpage

 \begin{figure*}
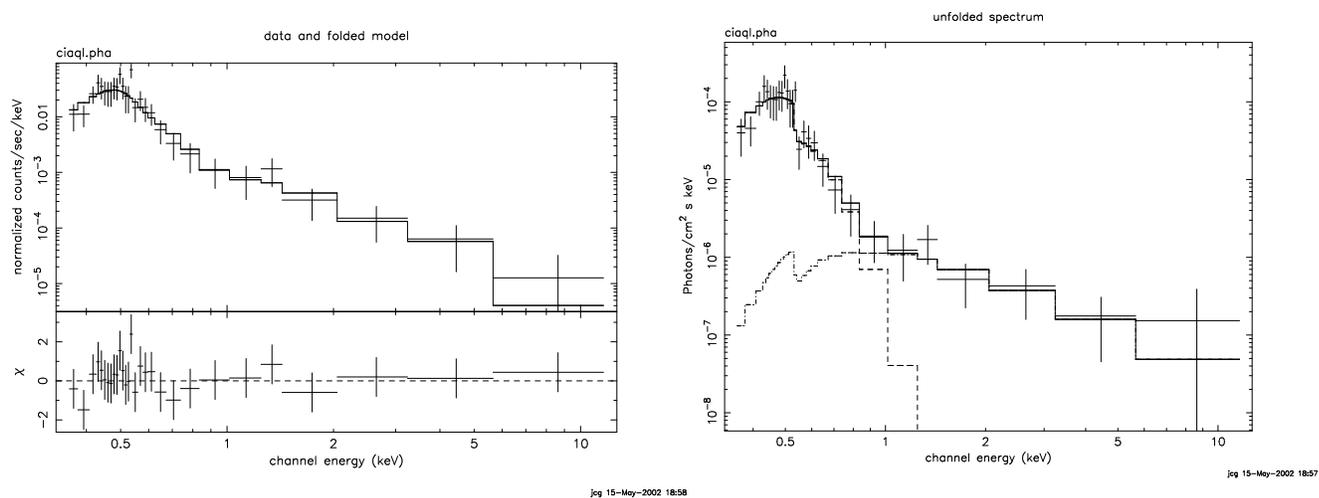

   \vbox{\psfig{figure=ciaql_bbpo_csp.ps,width=9cm,angle=270}}
  \vspace*{-6.5cm}\hspace*{8.8cm}
   \vbox{\psfig{figure=ciaql_bbpo.ps,width=8.5cm,angle=270}}
   \vspace*{-0.2cm}
   \caption[xspec]{X-ray spectrum of CI Aql as measured (left panel) with
      \chan\ ACIS-S on August 1 2001, composed of a blackbody (dashed line
      in right panel) and a power law (dotted line in right panel) model.
   \label{bbpo}}
 \end{figure*}

 \begin{figure*}
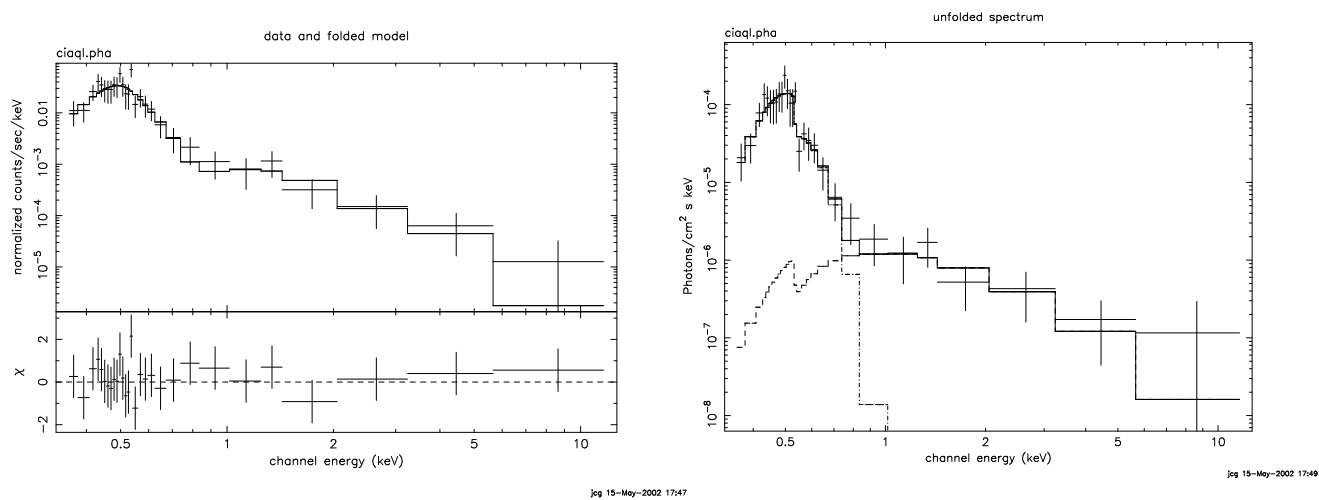

   \vbox{\psfig{figure=ciaql_1gauss+brems_csp.ps,width=9.cm,angle=270}}
  \vspace*{-6.5cm}\hspace*{8.8cm}
   \vbox{\psfig{figure=ciaql_1gauss+brems.ps,width=8.5cm,angle=270}}
   \vspace*{-0.2cm}
   \caption[xspec]{X-ray spectrum of CI Aql as measured with
      \chan\ ACIS-S on August 1 2001, composed of 1 Gaussian line
     (dash-dot line in right panel)
    and a thermal bremsstrahlung (dotted line in right panel) model.
   \label{gausbrems}}
 \end{figure*}

 \begin{figure*}
   \vbox{\psfig{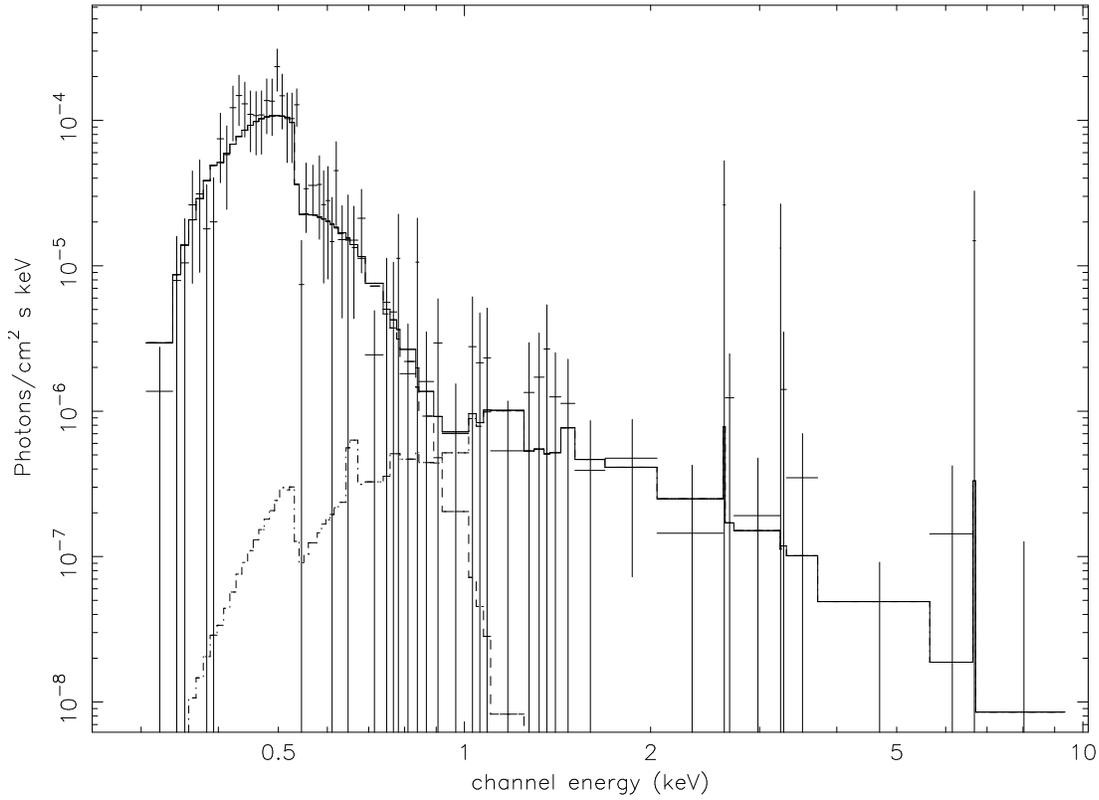}}\par
   \vspace*{-0.2cm}
   \caption[xspec]{X-ray spectrum of CI Aql as measured with
      \chan\ ACIS-S on August 1, 2001 composed of a blackbody and 
      a Raymond-Smith model. The binning has been reduced to only 1$\sigma$ 
     confidence to visualize the possible excesses at 1.3, 2.6, 3.3 and 
     6.6 keV (see text).
   \label{xsp4sig}}
 \end{figure*}

\end{document}